# Hydrogen-atom roaming reactions in water clusters: Unveiling an unusual dimension of water reactivity through first-principles calculations and machine learning

Rui Liu,[1] Baiqiang Liu,[1] Zhen Gong,[1] Zhaohua Cui,[1] Yue Feng,[1] and Zhigang Wang[1,2,*]

[1]*Key Laboratory of Material Simulation Methods & Software of Ministry of Education, College of Physics, Jilin University, Changchun 130012, China*

[2]*Institute of Theoretical Chemistry, College of Chemistry, Jilin University, Changchun 130023, China*

[*]*Email: wangzg@jlu.edu.cn (Z. W.)*

## Abstract

Water mediates a broad range of chemical reactions, including proton transfer, bond rearrangement, and conventional radical processes, defining a continuously expanding repertoire of intrinsic reactivity. However, roaming, a fundamental reaction mechanism that a departing fragment bypasses the minimum energy path to recombine, has not been identified in water itself. Here, we report the discovery of hydrogen-atom roaming reactions in water clusters. High-precision ab initio calculations of first-principles reveal that a neutral hydrogen atom departs as a radical, roams across the flat potential energy surface, and recombines along pathways that connect the same reactants and products as known hydrogen-bond network rearrangements. Interpretable machine learning analysis identifies the reactant dipole moment as the decisive switch governing whether roaming occurs, underpinned by exchange-repulsion and electrostatic interactions. Once roaming is initiated, polarizability and spin population determine barrier heights, while the charge distribution of the roaming hydrogen atom governs barrier widths, collectively shaped by electrostatic, orbital, and dispersion contributions. These findings establish hydrogen-atom roaming as a previously unrecognized intrinsic reaction class in water, complementing a fundamental dimension to the mechanistic picture of water reactivity.

**Introduction**

Water is one of the most fundamental substances in nature, providing the solvent environment essential for protein folding, enzyme catalysis, and DNA replication.[1-7] Beyond its role as a passive solvent, water participates directly in a wide range of reactions across a well-defined repertoire of intrinsic reaction classes, including proton transport, acid–base reactions, hydrolysis, redox, and conventional radical processes, many of which a proceed through hydrogen-bond network rearrangement.[8-14] Each newly identified reaction class within this repertoire has enriched the mechanistic picture of water reactivity, and to date, nearly every fundamental reaction mechanism has found its counterpart in water. One notable exception is roaming, a mechanism that strays from the minimum energy path along which conventional water reaction classes proceed.

The early roaming reaction was discovered in the photodissociation study of formaldehyde.[15] In such reactions, nearly detached fragments wander across a flat region of the potential energy surface before intramolecular abstraction to form molecular products.[16-22] Extensive studies have documented roaming across combustion, atmospheric, and interstellar chemistry, establishing it as a broadly significant and common reaction phenomenon.[23-27] Remarkably, water has emerged as an active participant in several of these processes. Prior studies have shown that ambient water molecules can modulate the roaming pathways of nitrobenzene,[28] and that a water-assisted roaming mechanism plays an important role in Criegee intermediate reactions relevant to atmospheric chemistry, in which new water molecules are generated in the course of these reactions.[29-31] In every documented case, however, water acts as an external modulator rather than the roaming substrate itself. Whether water is intrinsically capable of roaming remains an open question, leaving a missing piece in the mechanistic picture of water reactivity.

Resolving this question demands more than accurate energies and geometries for water systems, even obtained from high-precision ab initio calculations of first-principles. The intrinsic reactivity of water systems arises from multiple competing intermolecular interactions,[32,33] and the resulting high-dimensional complexity makes it difficult for conventional analysis to identify the key determinants that steer roaming. Machine learning (ML) offers a powerful route to this complexity, owing to its capacity to extract intricate, nonlinear relations from extensive molecular datasets.[34,35] Its application to water systems has shown success across a range of chemically distinct problems. For instance, classification and regression tree models have been applied to identify the local structural order parameters that govern water autoionization. This analysis revealed that specific anomalies in the hydrogen-bond network, rather than any single geometric criterion, serve as the proximal triggers for proton transfer.[36]

Moving beyond static descriptors, a decision tree trained on transition path-sampling data demonstrated that reactive and nonreactive trajectories in water can be distinguished with high accuracy, thereby exposing collective variables along the reaction coordinate that are inaccessible to conventional chemical intuition.[37] More recently, a combination of 23 structural and connectivity descriptors with ML regression models achieved quantitative prediction of the energetic stability of water clusters across a range of sizes, illuminating how hydrogen-bond topology evolves upon progressive aggregation.[38] These studies establish that descriptor-based ML approaches can decode the structural and reactive properties of water with near-quantitative accuracy. However, no such analysis has been directed at roaming. Crucially, this problem requires not only predictive accuracy but mechanistic transparency, a need met by interpretable approaches such as Shapley additive explanations (SHAP),[39] which quantify the contribution of individual molecular descriptors to specific outcomes.

Here, we report the discovery of hydrogen-atom roaming reactions in water clusters. First-principles calculations shows that a neutral hydrogen atom departs as a radical, roams across the flat potential energy surface, and recombines along pathways that connect the same reactants and products as known hydrogen-bond network rearrangements, with barrier heights comparable to the O-H bond dissociation energy. Interpretable machine learning analysis reveals the reactant dipole moment as the decisive switch controlling roaming occurrence, and further identifies the interaction mechanisms that govern barrier heights and widths. These findings establish hydrogen-atom roaming as a previously unrecognized intrinsic reaction class rooted in the intrinsic properties of water.

## Results and discussion

**Identification of hydrogen-atom roaming reactions and dataset construction**

A central question is whether roaming pathways exist alongside the known hydrogen-bond network rearrangement (HBNR) mechanisms in water clusters. We addressed this using the water dimer and trimer as prototype systems that encompass all elementary HBNR processes.[40-44] Conventional HBNR mechanisms have been extensively characterized for both the dimer, including acceptor tunneling, bifurcation tunneling, donor–acceptor interchange, and the trimer, including single flipping, bifurcation tunneling, concerted proton transfer.[45-47] Notably, analogous HBNR processes extend to larger water clusters, where the elementary HBNR established in the dimer and trimer recur and combine into more complex concerted mechanisms.[48-52] Here, our calculations across varied intermolecular separations reveal multiple hydrogen-atom roaming pathways that connect the same reactants and products as these known HBNR processes. Fig. 1a illustrates two such pathways in the water dimer, including roaming-mediated donor-acceptor exchange and roaming-mediated bifurcation. Both pathways exhibit geometrically similar roaming saddle points with closest H···H distances of 2.16–2.32 Å, and differ principally in the timing of the O-H bond rotation between Step 1 and Step 2. The water trimer displays three distinct roaming-mediated HBNR pathways (Fig. 1b), including roaming-mediated proton transfer, single flipping, and bifurcation, with closest H···H distances ranging from 1.77 to 2.62 Å. The key mechanistic differences among these pathways arise from the sequence of bond cleavage and rotation: in the Step 1, the roaming-mediated proton transfer and single flipping pathways both cleave all donor O-H bonds, whereas the bifurcation pathway cleaves one. In the Step 2, the former two differ in the mode of hydrogen reattachment and which O-H bond rotates. Notably, roaming also occurs at intermolecular separations of 3.18–3.33 Å without inducing structural rearrangement, and persists at separations exceeding 5.85 Å, where the dissociated hydrogen atom roams between monomers rather than circulating within the cluster ring. This long-range behavior suggests that hydrogen-atom roaming is not a structural anomaly but rather an intrinsic feature of water accessible across a broad separation region.

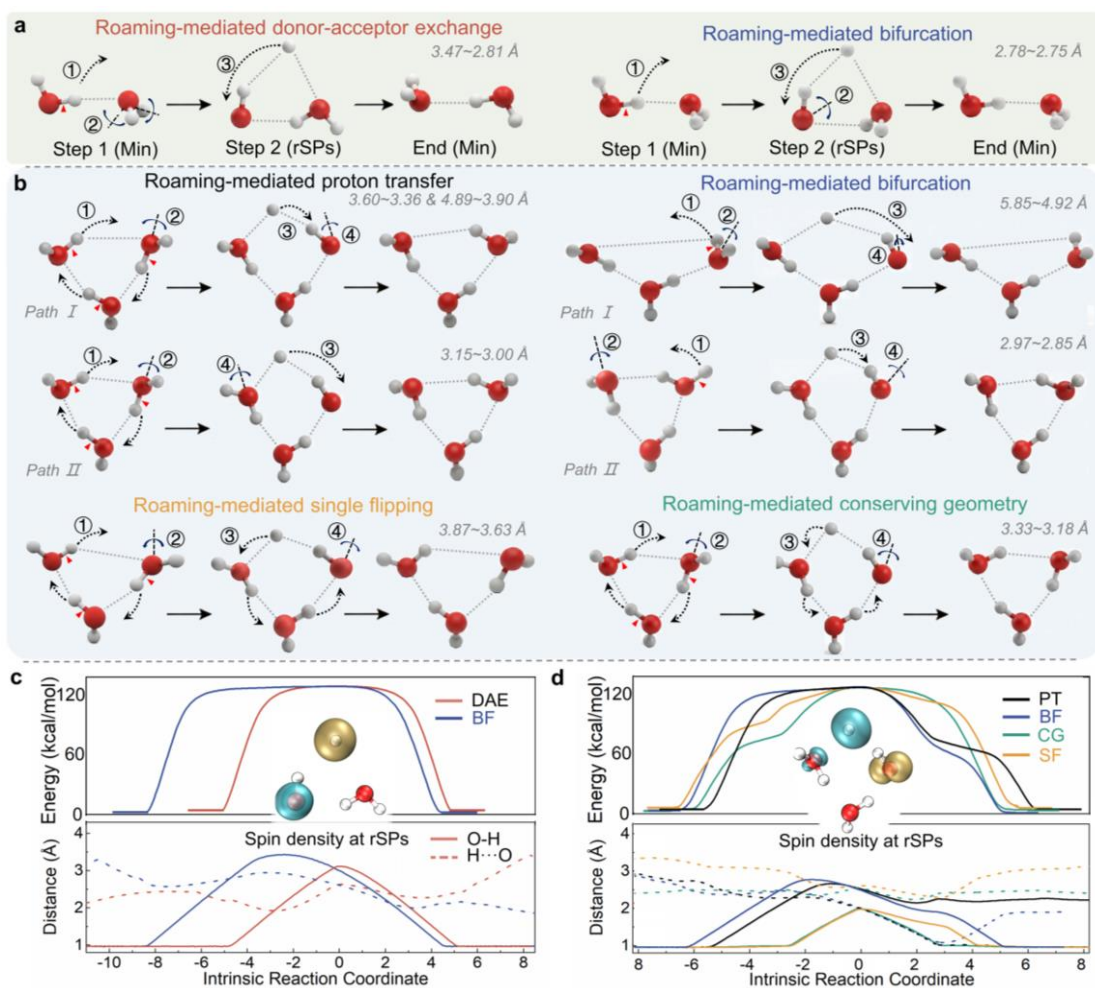

**Fig. 1 | Construction of the dataset of geometries and energies of hydrogen-atom roaming reactions in water dimer and trimer. a,** Roaming-mediated hydrogen-bond network rearrangement (HBNR) pathways in the water dimer. **b,** Roaming-mediated HBNR pathways in the water trimer. Dotted lines indicate the O-H axis (C2 axis). Each pathway proceeds through three stages: Step 1 (Min), Step 2 (rSPs), and End (Min). **c, d,** Potential energy profiles and variation in O-H and H···O distances along the intrinsic reaction coordinate for representative roaming-mediated processes in the dimer (c) and trimer (d). The origin (x = 0) corresponds to the roaming saddle point. Inset panels show spin populations at the roaming saddle points. Electronic energies were computed at the spin-unrestricted CCSD(T)/AVTZ level.

**Table 1. Detailed definitions of the 15 features screened as Input X in machine learning model.**

| Feature | Detailed definition | Feature | Detailed definition | Feature | Detailed definition |
|---------|---------------------|---------|---------------------|---------|---------------------|
| DM | Dipole moment | OHO | O-H···O angle | E | Interaction energy |
| Gap | HOMO-LUMO gap | OH | O-H bond length | $E^{elec}$ | Electrostatic energy |
| Pol | Polarizability | HO | H···O distance | $E^{exch}$ | Exchange-repulsion energy |
| Chg | Charge | OO | O···O distance | $E^{orb}$ | Orbital energy |
| SP | Spin population | $E^{tot}$ | Total energy | $E^{disp}$ | Dispersion energy |

To clarify the mechanistic nature of roaming, we analyzed potential energy surface, geometric changes, and spin populations at representative roaming saddle points (Fig. 1c, d). All roaming pathways exhibit relative barrier heights of ~122 kcal/mol, comparable to the O-H bond dissociation energy of ~118 kcal/mol,[53,54] consistent with such reactions occurring in the near-dissociative regions. The barrier widths, however, vary substantially across different roaming types. All roaming saddle points retain a highly spin-polarized singlet state with similar spin population patterns (Supplementary Fig. 1), consistent with the radical character of the roaming hydrogen atom. Interaction region indicator isosurfaces, localized orbital locator analysis, and energy decomposition trends (Supplementary Fig. 2) consistently indicate that roaming occurs within loosely bound van der Waals regions,[55-57] in contrast to the covalent character of conventional transition states. Having identified these pathways, we next constructed a comprehensive dataset to explore the physical factors that influence when and how roaming occurs.

The dataset is organized into input descriptors (Input X) and target labels (Output Y) so that the models can learn quantitative relations between molecular descriptors and roaming behaviors. Output Y addresses two tasks, including classification of roaming versus non-roaming classes, and regression of the factors controlling barrier height and width of these reactions. Input X comprises a set of physically meaningful descriptors that capture the electronic, geometric, and energetic properties (Table 1). Electronic descriptors include the dipole moment (DM), HOMO-LUMO gap (Gap), polarizability (Pol), atomic charges (Chg), and spin populations (SP). Geometric descriptors that characterize hydrogen bonding strength and directionality include O–H···O angle (OHO), O-H bond length (OH), H···O distance (HO), and O···O distance (OO). Energetic descriptors include the total energy ($E^{tot}$), interaction energy (E), electrostatic ($E^{elec}$), exchange-repulsion ($E^{exch}$), orbital ($E^{orb}$), and dispersion ($E^{disp}$) components.

**Model establishment and evaluation**

The machine learning workflow used to capture nonlinear descriptor relations underlying roaming behavior is summarized in Fig. 2. The curated dataset encodes geometric, electronic, and energetic descriptors for tens of thousands of the configurations, providing a foundation for model development. A classification model is first trained to distinguish roaming from non-roaming classes, with predictive performance evaluated using a random-extracted and recoverable cross-validation (RE-RCV) protocol,[58,59] which offers improved stability over the conventional k-fold validation for datasets of this complexity. Complementary regression models are then built to quantify how the individual descriptors govern roaming barrier height and width, with performance assessed by mean absolute error (MAE), mean squared error (MSE), root mean squared error (RMSE), and coefficient of determination ($R^2$). This workflow integrates the dataset construction, model establishment, model evaluation, and feature analysis into a unified, interpretable framework for exploring the factors that control roaming reactions in water clusters.

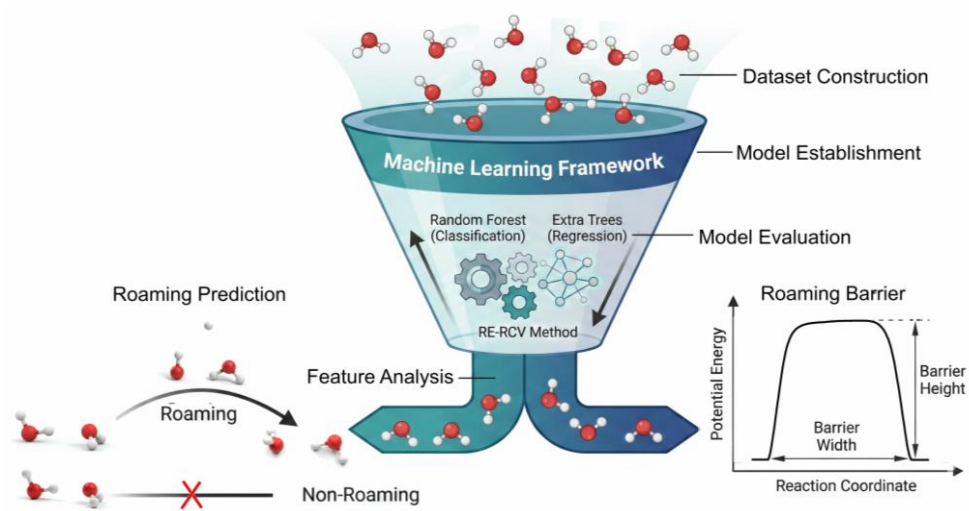

**Fig. 2 | Interpretable machine learning framework incorporating dataset construction, model establishment, model evaluation, and feature analysis.** In the RE-RCV evaluation protocol, a randomly extracted test set is used for evaluation and subsequently returned to the full dataset before the next extraction cycle.

Model evaluation is central to selecting an algorithmic structure that captures roaming behavior. Five representative classification algorithms were examined, including the support vector machine (SVM),[62] neural-network model (NNM),[63] random forest (RF),[64] k-nearest neighbor (KNN),[65] and naive Bayes (NB).[66] Each algorithm embodies a distinct inductive bias and therefore probes distinct aspects of the descriptor space. Although k-fold cross-validation is widely used, it can yield unstable accuracy estimates when dataset sizes are limited. To ensure robust evaluation, we employed the RE-RCV protocol, in which 20% of the of importance, and the red-to-blue color scale represents high-to-low feature values. **g,** SHAP contribution differences for each descriptor toward roaming versus non-roaming classification. **h,** Mean absolute SHAP values for each descriptor across roaming and non-roaming classes. **i,** Energy decomposition contributions to pathway classification, showing the relative importance of the four interaction energy components.

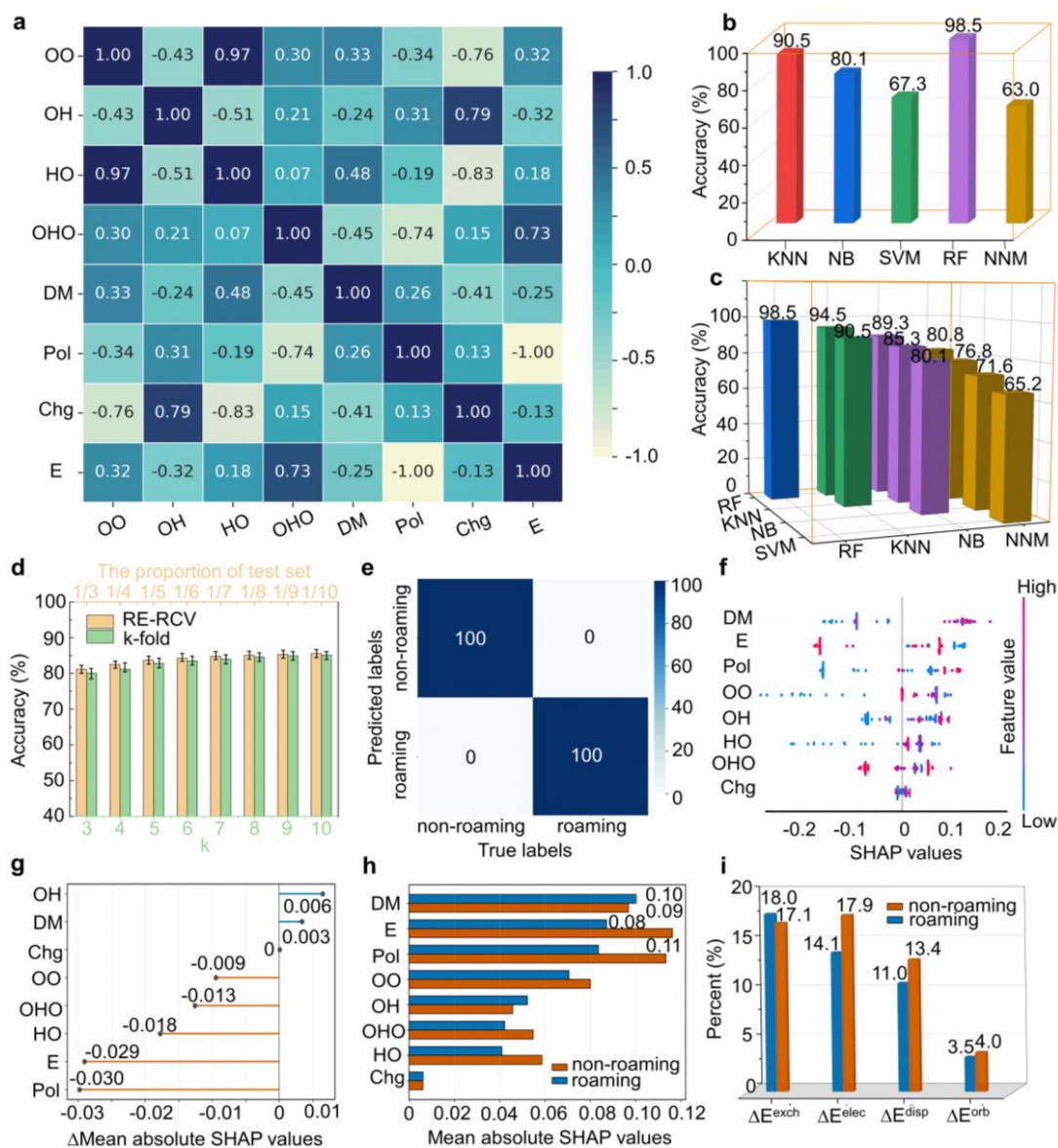

**Fig. 3 | Machine learning classification of roaming versus non-roaming cases in water dimer and trimer. a,** Pearson correlation matrix among the eight screened descriptors. Values within each cell are Pearson correlation coefficients; positive and negative values indicate positive and negative correlations, respectively. Classification accuracies for **b,** single ML models and **c,** stacked ML models evaluated by RE-RCV. **d,** Accuracy and associated error range for the RF model evaluated by RE-RCV and k-fold cross-validation, each repeated 100 times. **e**, Confusion matrix of the final RF model. The values in each cell represent model precision on a 0–100% color scale. **f,** Global SHAP feature importance ranking for roaming and non-roaming classification. Features are ranked in descending order

the data are randomly drawn as a test set, accuracy is recorded, and the extracted data are returned to the full dataset. This process is repeated this cycle 100 times, yielding statistically reliable accuracy estimates that reflect the intrinsic performance of each model. RE-RCV evaluation exposes pronounced performance differences among single model classifiers (Fig. 3b). The RF model achieves the highest performance, with an accuracy of 98.5%. The KNN model also performs well, with an accuracy above 90%. NB model reaches a moderate accuracy level, while SVM and NNM show noticeably lower performance. These results suggest that ensemble-based and distance-based classifiers are better suited to capturing the nonlinear descriptor relations associated with roaming reactions than margin-based or neural-network models. Given that limited datasets can render individual algorithms susceptible to overfitting and biased decision boundaries,[67] we constructed stacked models by pairing algorithms to reduce these issues and produce a unified predictor.

The stacked models for roaming pathway classification in water clusters (Fig. 3c) further confirm the dominant role of RF model. The model constructed from two RF classifiers performs equivalently to the single RF model, demonstrating that RF has captured the dominant predictive patterns and that additional stacking provides no further improvement. Combinations of RF and KNN also perform well, with the accuracies approaching 90%. These results confirm that the RF model offers the most robust classification performance among the algorithms tested.

To assess the reliability of the evaluation strategy, we directly compared RE-RCV and k-fold cross-validation using the RF classifier (Fig. 3d). Both methods were repeated 100 times with continuously varied test sets to quantify the sensitivity of the reported accuracy to data partitioning. The resulting accuracy distributions show that RE-RCV produces a substantially narrower error range than k-fold, confirming its superior stability, as further supported by the consistently lower coefficient of variation observed for RE-RCV across all models.

In addition to the overall model accuracy, it is equally important to assess the prediction reliability within each Output Y class. This class-specific prediction probability, referred to as model precision, reflects how often the predicted class matches the true class and therefore provides a more reliable measure of classification performance. The confusion matrix of the final model (Fig. 3e) shows precision of approximately 100% for both roaming and non-roaming classes. This balanced performance indicates that the classifier does not favor one class over the other and classifies roaming and non-roaming classes with comparable reliability across the dataset.

**Feature analysis**

To understand which descriptors drive the roaming versus non-roaming classification, we applied SHAP analysis to the final classifier. The global SHAP ranking (Fig. 3f) reveals a clear, physically interpretable hierarchy among the eight screened features. Electronic response properties, including the reactant dipole moment, total energy, and polarizability, emerge sequentially as the most influential descriptors, suggesting that the electronic properties of the reactant play a dominant role in determining whether a roaming pathway is accessible. Geometric descriptors contribute more moderately but provide additional information about the spatial arrangement of the roaming hydrogen atom.

Class-resolved SHAP analysis further clarifies how descriptor importance shifts between pathway types (Fig. 3g, h). Roaming pathways show enhanced sensitivity to O-H bond elongation, whereas descriptors associated with the broader hydrogen-bond network contribute more to non-roaming cases. This redistribution suggests that roaming becomes accessible when the system samples a weakly and broad separation regime. In contrast, electronic descriptors remain consistently influential across both classes. These results indicate that roaming is not triggered by a single parameter but instead reflects a cooperative interplay among multiple descriptors. The cooperative effect of hydrogen bonds is known to be sensitive to geometric boundaries,[68,69] providing a structural basis for the multi-descriptor dependence of roaming pathway selection.

Complementary interaction analysis (Fig. 3i) quantifies the contributions of intermolecular energy terms to pathway classification. Exchange-repulsion and electrostatic components jointly account for approximately 67% of the total contribution, while dispersion contributes nearly 24%. This partitioning supports a picture in which pathway selection is governed by a cooperative balance among exchange-repulsion, electrostatics, and dispersion, consistent with the SHAP feature analysis.

**Regression models for roaming barrier height and width**

To identify the descriptors governing roaming barrier height and width, we trained regression models on the same geometric, electronic, and energetic dataset. Model performance was assessed via MAE, MSE, RMSE, $R^2$, and deviation metrics (Supplementary Figs. 3-10). Among all tested algorithms, the ExtraTrees regression (ET)[70] delivered the highest predictive accuracy, followed by random forest (RF),[64] CatBoost (CB),[71] gradient boosting (GB),[72] linear regression (LG),[73] XGBoost (XB),[74] ridge regression (Ridge),[75] and partial least-squares regression (PLSR).[76] The superior performance of nonlinear ensemble methods underscores their capacity to capture complex, nonadditive relations between descriptors and barrier properties.

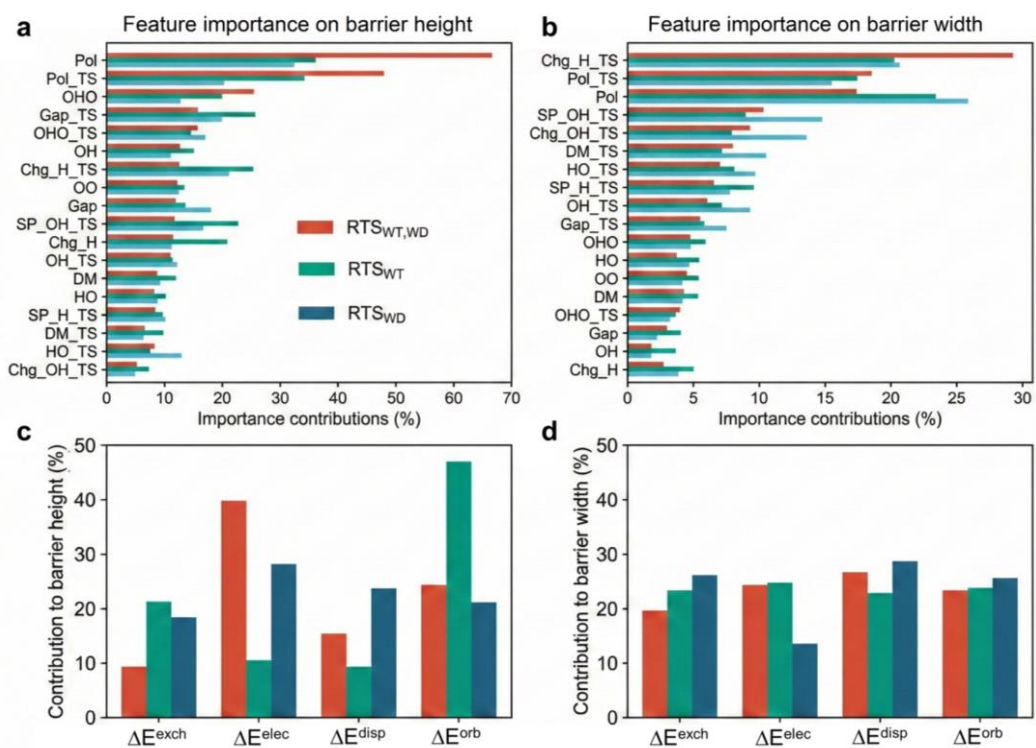

**Fig. 4 | Machine learning analysis of roaming barrier height and width in water dimers and trimers and their underlying physical determinants. a,b,** Feature importance profiles for barrier height (a) and barrier width (b), with descriptors ranked in descending order of importance. **c,d,** Energy decomposition contributions governing barrier height (c) and barrier width (d) for roaming saddle points of the water dimer ($RTS_{WD}$), water trimer ($RTS_{WT}$), and their combination ($RTS_{WD,WT}$). All regression models employed the ExtraTrees regression with 100-fold cross-validation.

Using the ET model as the optimal regressor, we quantified feature importance for the water dimer, water trimer, and combined datasets (Fig. 4a). For barrier height in combined dataset, the polarizabilities of the reactant and of the roaming saddle point dominate, together accounting for approximately 83% of total importance. This result suggests that electronic flexibility, that is, the capacity of the electron density to reorganize in response to structural deformation is closely linked to the barrier height of roaming reactions. When dimer and trimer systems are analyzed separately, the dimer follows the combined trend with polarizability dominant, whereas the trimer exhibits a more distributed importance profile in which the O-H spin population also emerges as a significant contributor, indicating that spin redistribution plays an important role in larger clusters.

The determinants of barrier width follow a different pattern (Fig. 4b). In the combined dataset, the charge on the roaming hydrogen atom and polarizability together account for nearly 37.7% of total importance. As analyzed separately, water dimer shows polarizability and O-H spin population jointly accounting for approximately 69.4% of the importance, while the water trimer exhibits a more distributed profile in which the dipole moment and polarizability emerge as prominent contributors. These system-dependent differences indicate that the determinants of barrier width are more sensitive to cluster size than those governing barrier height, suggesting that as the hydrogen-bond network grows in complexity, collective electronic effects become increasingly important in determining the spatial extent of the roaming barrier.

To connect the ML feature importance results with the underlying intermolecular energetics, we performed energy decomposition analysis (Fig. 4c, d). For barrier height, electrostatic interactions dominate, with a non-negligible orbital contribution, together accounting for approximately 65% of the total. In contrast, the barrier width is dominated by electrostatic and dispersion interactions, accounting for nearly 45% of the total. These energy decomposition trends corroborate the ML feature importance results, providing a physically interpretable picture of the physical determinants dictating roaming barriers in water clusters.

## Conclusion

In summary, this work presents hydrogen-atom roaming reactions in water clusters. First-principles calculations indicate that a neutral hydrogen atom departs as a radical, roams across a flat potential energy surface, and recombines to yield the same reactants and products as conventional hydrogen-bond network rearrangements. The associated barrier heights, comparable to the O-H bond dissociation energy, place these pathways in the near-dissociative region and thereby establish hydrogen-atom roaming as a previously unrecognized intrinsic reaction class in water. Interpretable machine learning analysis reveals the physical origins governing these roaming pathways. The reactant dipole moment emerges as a decisive switch that controls whether roaming occurs, underpinned by exchange-repulsion and electrostatic interactions. Once roaming is initiated, polarizability and spin population determine barrier heights, while the charge distribution of the roaming hydrogen atom governs barrier widths, collectively shaped by electrostatic, orbital, and dispersion contributions. These findings suggest that hydrogen-atom roaming constitutes a fundamental dimension rooted in the intrinsic properties of water itself, thereby complementing the mechanistic picture of water reactivity.

## Method

**Machine learning details**

For the classification task, five supervised learning algorithms were examined to distinguish roaming from non-roaming classes, including support vector machine (SVM),[62] neural-network model (NNM),[63] random forest (RF),[64] k-nearest neighbor (KNN),[65] and naive Bayes (NB).[66]

For the regression task, roaming barrier height and width were predicted using a suite of regression algorithms, including the ExtraTrees regression (ET),[70] random forest (RF),[64] CatBoost (CB),[71] gradient boosting (GB),[72] linear regression (LG),[73] XGBoost (XB),[74] ridge regression (Ridge),[75] and partial least-squares regression (PLSR).[76]

To identify and remove redundant descriptors, pairwise linear correlations among all features were quantified using Pearson's correlation coefficient, which normalizes the covariance between two variables by their individual standard deviations (Eq. 1). Feature pairs with $|p| > 0.8$ were considered strongly collinear; only one descriptor from each such pair was retained to ensure the final feature set remained compact and non-degenerate,

$$p = \frac{\sum_{i=1}^{n}[(x_i - \bar{x}) \times (y_i - \bar{y})]}{\sqrt{\sum_{i=1}^{n}(x_i - \bar{x})^2} \times \sqrt{\sum_{i=1}^{n}(y_i - \bar{y})^2}} \qquad (1)$$

where $x_i$ and $y_i$ are the values of two features in the respective data, $\bar{x}$ and $\bar{y}$ are the average values of these two features.

**Model evaluation method**

To evaluate model performance with enhanced statistical robustness, a random-extracted and recoverable cross-validation (RE-RCV) strategy was employed. In each iteration, a fixed proportion (20%) of samples was randomly drawn from the full dataset to serve as a test set, and the model was trained on the remaining data. After the accuracy for that iteration was recorded, the extracted samples were returned to the full dataset. This process was repeated 100 independent times, with each test subset newly and independently drawn. Final model accuracy was obtained by averaging across all 100 iterations.

**Computational details**

All stationary point descriptors, including DM, Gap, Pol, Chg, SP, OHO, OH, HO, OO, $E^{tot}$, E, $E^{elec}$, $E^{exch}$, $E^{orb}$, and $E^{disp}$, were computed using first-principles calculations at the spin-unrestricted CCSD(T)//B3LYP/AVTZ level of theory. Highly spin-polarized roaming saddle points were located using broken-symmetry DFT with transition-state searches based on the geometry direct inversion in the iterative subspace method.[77] Intrinsic reaction coordinate calculations employing the Local Quadratic Approximation method[78] were performed along the spin-polarized potential energy surface to confirm the connectivity between roaming saddle points and their corresponding reactant and product minima. All electronic structure calculations were carried out using Gaussian 09.[79] Energy decomposition analysis was performed using the sobEDA code.[80] Electronic structure and property analyses were conducted using Multiwfn.[81]

# Data availability

The data that support the findings of this study are available from the corresponding author upon reasonable request.

## Acknowledgements

The authors wish to acknowledge Dr. Xinrui Yang and Ms. Siyang Liu for academic discussions. This work was supported by the Science and Technology Development Program of Jilin Province of China (20250102014JC) and the National Key Research and Development Program of China (No. 2024YFA1409900). Z. W. also acknowledges the assistance of the

## Author contributions

Zhigang Wang initiated, designed and supervised the work. Rui Liu performed the theoretical simulations. Rui Liu, Baiqiang Liu, Zhen Gong, Zhaohua Cui, Yue Feng, and Zhigang Wang discussed the results and wrote the article.

## Competing interests

All authors declare no competing interests.